\documentclass[prl,twocolumn,floatfix,superscriptaddress]{revtex4-1}
\usepackage{dcolumn,amsmath}
\usepackage{graphicx}
\usepackage{bm}
\usepackage{hyperref}
\usepackage{color}
\usepackage{upgreek}

\usepackage[normalem]{ulem} 

\begin{document}

\title{New Nuclear Magnetic Moment of $^{209}$Bi - Resolving the Bismuth Hyperfine Puzzle}

\author{Leonid V. Skripnikov}\email{leonidos239@gmail.com}
\affiliation{Department of Physics, St. Petersburg State University, Universitetskaya 7/9, 199034 St. Petersburg, Russia}
\affiliation{Petersburg Nuclear Physics Institute named by B.P. Konstantinov of National Research Centre ``Kurchatov Institute'', 188300 Gatchina, Leningrad District, Russia}
\author{Stefan~Schmidt}
\affiliation{Institut f\"ur Kernphysik, Technische Universit\"at Darmstadt, 64289 Darmstadt, Germany}
\affiliation{Institut f\"ur Physik, Universit\"at Mainz, 55128 Mainz, Germany}
\author{Johannes~Ullmann}
\affiliation{Institut f\"ur Kernphysik, Technische Universit\"at Darmstadt, 64289 Darmstadt, Germany}
\affiliation{Institut f\"ur Kernphysik, Westf\"alische Wilhelms-Universit\"at M\"unster, 48149 Münster, Germany}
\author{Christopher~Geppert}
\affiliation{Institut f\"ur Kernchemie, Universit\"at Mainz, 55128 Mainz, Germany}
\author{Florian~Kraus}
\affiliation{Anorganische Chemie und Fluorchemie, Philipps-Universit\"at Marburg, 35032 Marburg, Germany}
\author{Benjamin~Kresse}
\affiliation{Institut f\"ur Festk\"orperphysik, Technische Universit\"at Darmstadt, Hochschulstr. 6, 64289 Darmstadt, Germany}
\author{Wilfried~N\"ortersh\"auser}
\affiliation{Institut f\"ur Kernphysik, Technische Universit\"at Darmstadt, 64289 Darmstadt, Germany}
\author{Alexei~F.~Privalov}
\affiliation{Institut f\"ur Festk\"orperphysik, Technische Universit\"at Darmstadt, Hochschulstr. 6, 64289 Darmstadt, Germany}
\author{Benjamin~Scheibe} 
\affiliation{Anorganische Chemie und Fluorchemie, Philipps-Universit\"at Marburg, 35032 Marburg, Germany}
\author{Vladimir~M.~Shabaev}
\affiliation{Department of Physics, St. Petersburg State University, Universitetskaya 7/9, 199034 St. Petersburg, Russia}
\author{Michael~Vogel}
\affiliation{Institut f\"ur Festk\"orperphysik, Technische Universit\"at Darmstadt, Hochschulstr. 6, 64289 Darmstadt, Germany}
\author{Andrey~V.~Volotka}
\affiliation{Helmholtz-Institut Jena, D-07743 Jena, Germany}
\affiliation{Department of Physics, St. Petersburg State University, Universitetskaya 7/9, 199034 St. Petersburg, Russia}
\date{07.03.2018}
\begin{abstract}

A recent measurement of the hyperfine splitting in the ground state of Li-like $^{209}$Bi$^{80+}$ has established a ``hyperfine puzzle'' -- the experimental result exhibits a 7$\sigma$ deviation from the theoretical prediction [J. Ullmann et al., Nat. Commun. 8, 15484 (2017); J. P. Karr, Nat. Phys. 13, 533 (2017)].
We provide evidence that the discrepancy is caused by an inaccurate value of the tabulated nuclear magnetic moment ($\mu_I$) of $^{209}$Bi. We perform relativistic density functional theory and relativistic coupled cluster calculations of the shielding constant that should be used to extract the value of $\mu_I(^{209}{\rm Bi})$ and combine it with nuclear magnetic resonance measurements of Bi(NO$_3$)$_3$ in nitric acid solutions and of the hexafluoridobismuthate(V) BiF$_6^-$ ion in acetonitrile. The result clearly reveals that $\mu_I(^{209}{\rm Bi})$ is much smaller than the tabulated value used previously. Applying the new magnetic moment shifts the theoretical prediction into agreement with experiment and resolves the hyperfine puzzle.
\end{abstract}
\maketitle

\section{Introduction}

A combined measurement of the hyperfine structure (HFS) splittings in hydrogenlike and lithiumlike ions of $^{209}$Bi has been suggested as early as 2001 \cite{Shabaev:01a} to be a sensitive probe for bound-state strong-field QED in the strongest static magnetic fields available in the laboratory. Such fields exist in the surrounding of heavy nuclei with nuclear spin and a large nuclear magnetic moment. The electron in H-like $^{209}$Bi$^{82+}$, for example, experiences on average a magnetic field of about 30\,000\,T, more than 1000 times stronger than available with the strongest superconducting magnet.    
%
According to \cite{Shabaev:01a}, a special combination of the ground-state HFS splittings in H-like and Li-like ions ($\Delta E^{(1s)}$ and $\Delta E^{(2s)}$, respectively) of the same nuclear species, called the specific difference 
%
%
%
\begin{eqnarray}
\Delta 'E = \Delta E^{(2s)} - \xi \Delta E^{(1s)},
\end{eqnarray}
provides the best means to test bound-state strong-field QED in the magnetic regime.
Here, the parameter $\xi = 0.16886$ \cite{Shabaev:01a,Volotka:12} is chosen to cancel the contributions of the nuclear-magnetization distribution  (Bohr-Weisskopf effect) to $\Delta E^{(1s)}$ and $\Delta E^{(2s)}$.
This is required since the uncertainties of these contributions to the HFS splittings are commonly larger than the complete QED contribution and have failed all previous attempts to perform a QED test solely based on the HFS splitting in H-like heavy ions. However, at the time of the proposal \cite{Shabaev:01a} the experimental uncertainty of the HFS splitting in the Li-like $^{209}$Bi$^{80+}$ extracted 
from x-ray emission spectra \cite{Beiersdorfer:98} was far too high to verify the predictions
for $\Delta 'E$.
The first laser spectroscopic observation of the splitting reported in 2014 was orders of magnitude more precise but still limited by systematical uncertainties \cite{Lochmann:14}. Finally, further improvement in accuracy by more than an order of magnitude was recently reported \cite{Ullmann:15,Ullmann:17} but the result was surprisingly more than $7\sigma$ off from the latest theoretical prediction \cite{Volotka:12}. 
Since the experimental nuclear magnetic moment $\mu_I$ of $^{209}$Bi enters the calculation of the specific difference, an incorrect value will lead to a proportional change in $\Delta^\prime E$, which could be responsible for the discrepancy \cite{Karr:17}.
We also note that in Ref. \cite{Urrutia:96} the discrepancy between theory and experiment on the HFS splitting in H-like Ho was ascribed to an inaccurate value of the nuclear magnetic moment of ${^{165}}$Ho.

We have reexamined the literature value $\mu_I$($^{209}$Bi) 
obtained from nuclear magnetic resonance (NMR) experiments from a theoretical point of view. This has motivated new NMR measurements of bismuth ions in different chemical environments. Results of these experiments  are reported and analyzed applying high-level 
four-component relativistic coupled cluster theory for advanced chemical shift calculations. We show that our result can completely resolve the hyperfine puzzle established in \cite{Ullmann:17}. 
The specific difference $\Delta ' E$ has, so far, always been calculated using 
the magnetic moment $\mu_I(^{209}{\rm Bi})=4.1106(2)\mu_N$ tabulated in
\cite{Raghavan:89}. 
This value was obtained using the uncorrected (for shielding effects) experimental value of the magnetic moment $\mu_I(^{209}{\rm Bi})=4.03910(19)\mu_N$ reported in an NMR study \cite{Ting:53} of bismuth nitrate, Bi(NO$_3$)$_3$, which was then  combined with the shielding constant for the Bi$^{3+}$ cation calculated in \cite{Johnson:68}.
In \cite{Bastug:96} the self-consistent relativistic molecular Dirac-Fock-Slater calculation of the shielding constant of the Bi(NO$_3$)$_3$ molecule using the Lamb formula \cite{Lamb:41} was performed.
The final value, $\sigma=17290(60)$\,ppm, with very small uncertainty was obtained by combining relativistic random phase approximation calculation of the Bi$^{3+}$ cation (17270 ppm) with the molecular correction. The authors concluded that the molecular correction is very small and thus supported the value from \cite{Raghavan:89}. 

However, 
the authors of \cite{Bastug:96} have not taken into account chemical processes that occur in an aqueous solution of bismuth nitrate molecule Bi(NO$_3$)$_3 \cdot$5H$_2$O: the compound
dissociates and the Bi$^{3+}$ cation is surrounded by water molecules (hydration). Neither the completeness nor the exact form of hydration as a function of concentration, \textit{p}H or temperature is well understood.  
While it was suggested in \cite{Fedorov:98} that in strongly acidic solutions Bi$^{3+}$ exists as hexaaquabismuth(III)-cation [Bi(H$_2$O)$_6$]$^{3+}$, more recent studies \cite{Naslund:00} expect that the hydrated form is rather [Bi(H$_2$O)$_8$]$^{3+}$. We found that in both cases the electronic structure of the $n$-coordinated complex significantly differs from the Bi(NO$_3$)$_3$ molecule considered in \cite{Bastug:96}, which is expected. The molecular environment in 
Bi(III/V)-containing
complexes strongly contributes to the shielding constant and a considerable chemical shift is introduced. 
Consequently, the value of the shielding constant obtained in Ref.\,\cite{Bastug:96} cannot be used for the precise extraction of the $^{209}$Bi magnetic moment from the experimental NMR data. 

There is, however, additional NMR data for another Bi containing system: the
hexafluoridobismuthate(V) anion ($^{209}$BiF$_6^-$) \cite{Morgan:83}. It has seven atoms and high spatial symmetry.
According to Morgan \textit{et} al.\ \cite{Morgan:83}, a measurement of BiF$_6^-$ with reference to a saturated solution of bismuth nitrate in concentrated nitric acid gave a chemical shift of $-24$\,ppm. Unfortunately, there is an inconsistency in the reported experimental data of \cite{Morgan:83}, since the measured frequency ratio is given as $\nu(^{209} {\rm BiF}_6^-) / \nu(^1{\rm H})$ =0.16017649(10). The comparison of this ratio with the one reported in \cite{Ting:53} indicates a massive chemical shift of about $\delta \approx +3200$\,ppm instead of $-24$\,ppm. We have performed NMR measurements of both samples to clarify these discrepancies.  

\section{Experiment}
Since a dependence of the chemical state of the Bi$^{3+}$ ions in an aqueous solution is expected but details on the sample preparation are missing in the original NMR measurements \cite{Ting:53}, we performed a systematic study using various bismuth nitrate solutions.
Samples of ``Bi(NO$_3$)$_3$'' solutions were prepared with concentrations of 2.5\%, 5\% and 10\% Bi$^{3+}$ (wt \%) in concentrated (65 wt \%) and diluted aqueous solutions (50, 30, 20, 10 wt \%) of nitric acid (HNO$_3$). 

\begin{figure}[t]
\includegraphics[width=0.98\linewidth]{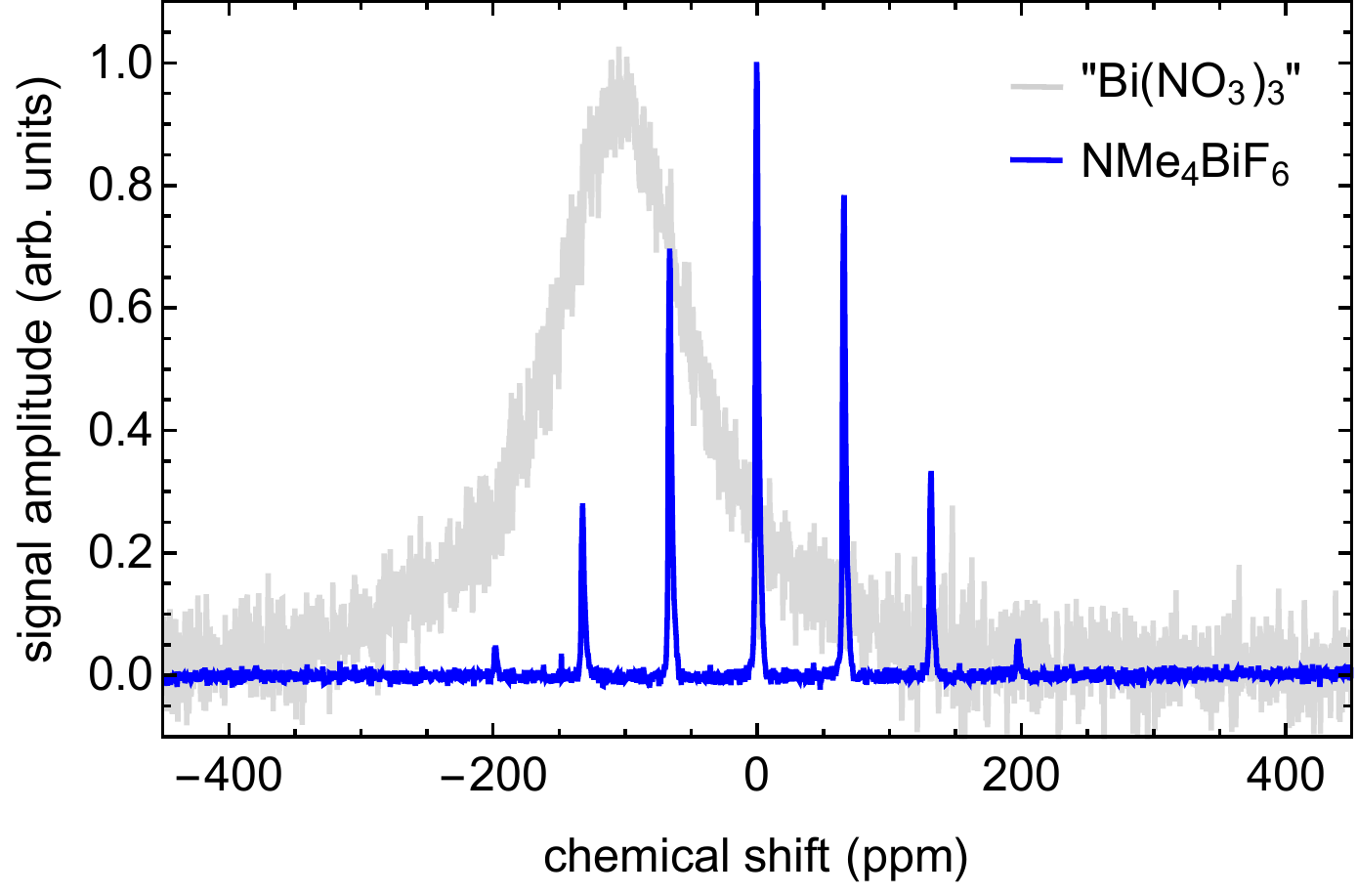}
\caption{\label{fig:Spectra} NMR spectra of Bi(NO$_3$)$_3$ solution (10\% Bi (wt \%)) in concentrated nitric acid (gray) and  NMe$_4$BiF$_6$ diluted in acetonitrile (blue).}
\end{figure}

BiF$_6^-$ anions were obtained by dissolution of $\mathrm{(CH_3)_4N^+BiF_6^-}$ (NMe$_4$BiF$_6$)
in acetonitrile to a saturated solution
\cite{Note1}. 

All NMR measurements were performed at an 8.4-T magnet using the same double resonance probe for $^{209}$Bi NMR and $^{1}$H NMR calibration with tetramethylsilane. The sample temperature was stabilized with an accuracy of 1\,K employing a constant gas flow tempered by an electric heater. Spectra were obtained from the free induction decay following a 90$^\circ$ pulse of 3.5\,$\upmu$s length for $^{209}$Bi. 

Typical spectra of the $^{209}$Bi atoms in BiF$_6^-$ and in the nitrate solution are shown in Fig.\,\ref{fig:Spectra}. The advantage of BiF$_6^-$ is obvious. It exhibits a much narrower linewidth (200 Hz) and the septet arising from indirect spin coupling of $^{19}$F atoms directly bonded to the bismuth atom assures the chemical environment.
The observed ratio of the peak intensities is close to the expected ratio 1\,:\,6\,:\,15\,:\,20\,:\,15\,:\,6\,:\,1 and a spin-spin coupling of 3807(14)\,Hz was determined, in good agreement with \cite{Morgan:83}.
Note that a $^{19}$F spectrum of the sample was taken as well and a decet consistent with the coupling of an $I=9/2$ nucleus to an octahedral environment of six fluorine atoms was observed. 
The signal from the nitrate solution is much wider. Even at the highest temperature of 360\,K, the width of the $^{209}$Bi spectra was 4.4\,kHz due to the short spin-lattice and spin-spin relaxation times of $\approx 70$\,$\upmu$s. This width limits the accuracy of the $^{209}$Bi resonance frequency in the solution of the nitrate to 1\,ppm. The chemical shift of Bi$^{3+}$ in the solution of the bismuth nitrate with respect to Bi$^{5+}$ in BiF$_6^-$ is $-106$\,ppm, larger than the $-24$\,ppm reported in \cite{Morgan:83}. Contrary to \cite{Flynn:59} we found that the variation of the bismuth concentration between mass fractions of 2\% and about 40\% (saturation) in nitric acid of 30\% had no appreciable effect on the measured Larmor frequency as long as temperature and nitric acid concentration were kept constant.

\begin{figure}[t]
\includegraphics[width=0.98\linewidth]{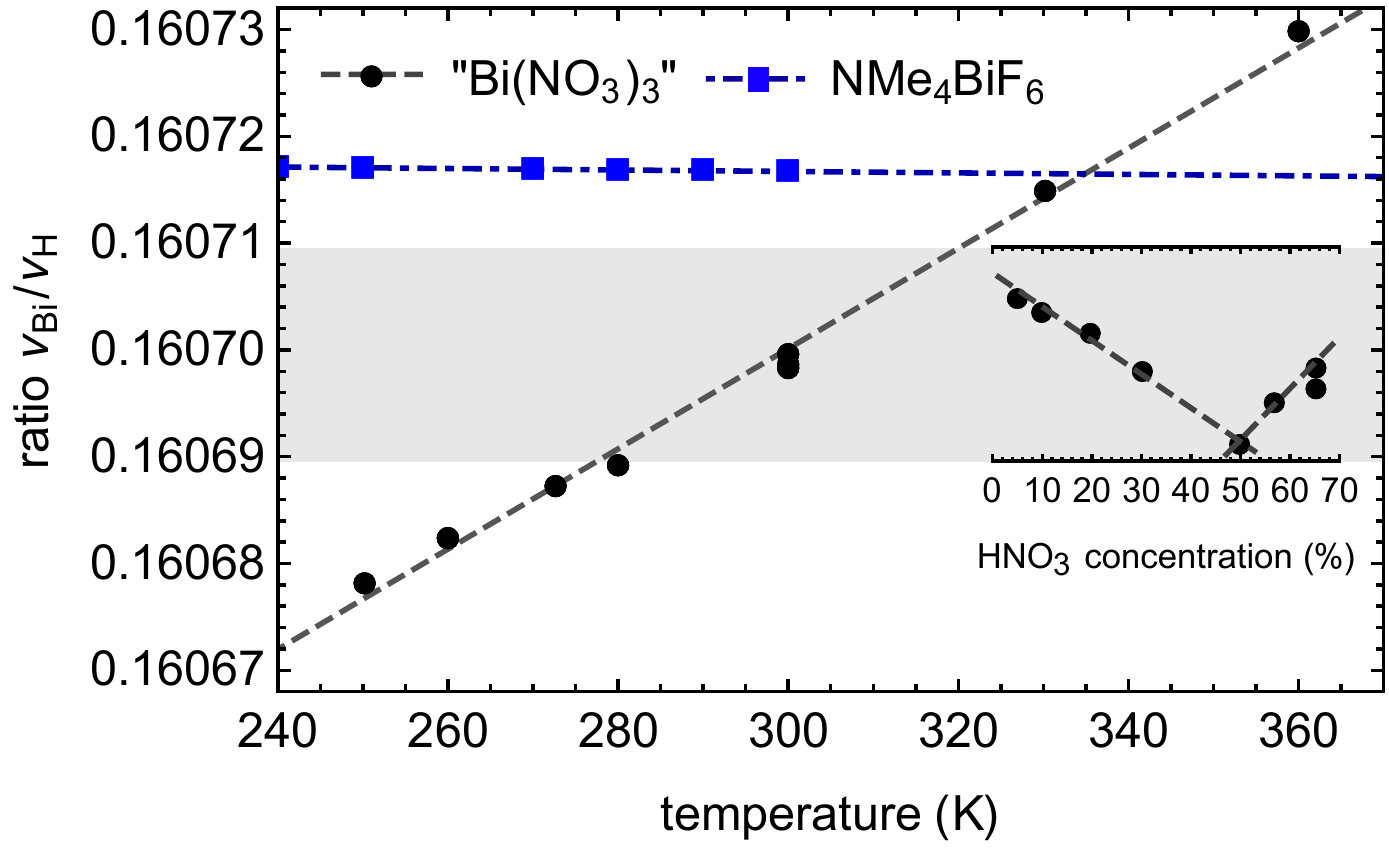}
\caption{\label{fig:TempDependence} Temperature and HNO$_3$-concentration dependency of the NMR Larmor-frequency ratios of bismuth and hydrogen. A strong temperature effect is observed for Bi(NO$_3$)$_3$ solutions, here exemplified for a 
10\% Bi$^{3+}$ (wt \%) solution in concentrated nitric acid (black), whereas only a minor effect was measured for NMe$_4$BiF$_6$ dissolved in acetonitrile (blue).
Inset: Larmor-frequency ratios 
measured by NMR in Bi(NO$_3$)$_3$ solutions with 2.5\% Bi$^{3+}$ (wt \%) in nitric acid (HNO$_3$) of various concentrations at 300\,K. 
The $y$ axis is identical to the main graph and the gray band represents the total variation.}
\end{figure}

Variations of the Bi(NO$_3$)$_3$ sample temperature from 250 to 360\,K were performed with the sample of 10\% Bi in concentrated nitric acid (65\%). We observed a strong linear temperature dependence of the frequency ratio in this range (Fig.\,\ref{fig:TempDependence}) with a slope of $+4.69(13)\times 10^{-7}$\,K$^{-1}$, corresponding to about 3\,ppm/K,
which might be caused by the change of density.
For standard NMR conditions at 298.15\,K a frequency ratio of $\nu_\mathrm{^{209}Bi^{3+}}/\nu_\mathrm{H}=0.160699(1)$ was determined, where the given uncertainty is purely statistical. This value is in excellent agreement with 0.160696(6) reported in \cite{Ting:53}. The temperature dependency of BiF$_6^-$ is 2 orders of magnitude smaller ($\approx 20$\,ppb/K) and of opposite sign. At 298.15\,K the frequency ratio to the proton is 0.1607167(2) far off from the value provided in \cite{Morgan:83}. However, the latter matches our value if one simply flips two digits [$0.160{\bf 17}65(1) \to 0.160{\bf 71}65(1)$].  

Finally, we have studied the resonance position of Bi(NO$_3$)$_3$ as a function of the nitric acid concentration
(inset in Fig.\,\ref{fig:TempDependence}).
A clear dependence on the acidity is observed for all Bi$^{3+}$ concentrations, covering a range of typically $\approx 60$\,ppm. 

In summary, the results clearly demonstrate that a large uncertainty is connected with the extraction of the magnetic moment of $^{209}$Bi from NMR measurements in aqueous solutions of Bi(NO$_3$)$_3$. The influence of the chemical environment was strongly underestimated in theory since the calculations performed to extract the chemical shift do neither account for the temperature nor for the concentration or acidity of the sample. In this respect, BiF$_6^-$ is a much better candidate to obtain a reliable value of the magnetic moment which will be substantiated now also from a  theoretical point of view.  

\section{Theory}

In the presence of the external uniform magnetic field \textbf{B} and nuclear magnetic moment $\mu_j$ of $j-$th atom in a molecule the corresponding Dirac-Coulomb Hamiltonian includes the following terms:
\begin{equation}
 \label{HB}
H_B={\rm \bf{B}}\cdot \frac{c}{2}(\bm{r}_G \times \bm{\alpha}),
\end{equation}
\begin{equation}
 \label{HHFS}
H_{\rm hyp}=\frac{1}{c} \sum_j \bm{\mu}_j\cdot \frac{(\bm{r}_j \times \bm{\alpha})}{r_j^3},
\end{equation}
where $\bm{r}_G = \bm{r} - \bm{R_G}$, $\bm{R_G}$ is the gauge origin, 
$\bm{r}_j=\bm{r} - \bm{R_j}$, $\bm{R_j}$ is the position of nucleus $j$, and $\bm{\alpha}$ are the Dirac matrices.

The chemical shielding tensor of the nucleus $j$ can be defined as a mixed derivative of the energy with respect to the nuclear magnetic moment and the strength of the magnetic field
\begin{equation}
 \label{SHIELDINGDer}
\left.\sigma^j_{a,b}=\frac{\partial^2E}{\partial\mu_{j,a}\partial B_b} \right|_{\bm{\mu}_j=0,{\rm \bf{B}}=0}.
\end{equation}
We are interested in its isotropic part.

In the one-electron case the shielding tensor (\ref{SHIELDINGDer}) can be calculated by the sum-over-states method within the second-order perturbation theory with perturbations (\ref{HB}) and (\ref{HHFS}). In the relativistic four-component approach the summation should include both positive and negative energy spectra \cite{Aucar:99}.
The part associated with positive energy is called the ``paramagnetic'' term while the part associated with negative energy states is called ``diamagnetic term'' though only their sum is gauge invariant \cite{Aucar:99}.

To avoid an ambiguity in calculations utilizing finite basis sets due to the choice of the gauge origin $\bm{R_G}$ one can use the so-called London atomic orbitals (LAOs) method (see e.g.\ \cite{Olejniczak:12,Ilias:13} for details).
In Refs.\,\cite{DIRAC15,Olejniczak:12,Ilias:13} the four-component density functional theory (DFT) using response technique and LAOs has been developed to calculate the shielding constant (\ref{SHIELDINGDer}).
To construct the atomic basis sets for the unperturbed Dirac-Coulomb Hamiltonian calculations one often uses the restricted kinetic balance (RKB) method. However, in the presence of the external magnetic fields the usual relation between the large and small component changes. In Ref.\,\cite{Olejniczak:12} the scheme of magnetic balance (MB) in conjunction with LAOs was proposed to take into account the modified coupling which is utilised below.

Most of the chemical shift calculations for heavy atom compounds are performed within the (relativistic) DFT. The drawback of the theory is that it is hard to control the uncertainty of the results as there is no systematic way of improving it. Even combinations with high-level nonrelativistic \textit{ab initio} wave-function-based calculations are also questionable in the case of heavy atom compounds. In Refs.\,\cite{Skripnikov:16b,Skripnikov:15a,Petrov:17b,Skripnikov:17c} it was shown that for such properties as the hyperfine structure constant and the molecular $g$ factor, the relativistic coupled cluster method gives the most accurate results if there are no multireference effects. Therefore, this method has been adopted here to control the uncertainty of the DFT results.

\section{Electronic structure calculation details}

In the present study we have used atomic basis sets of different qualities.
The NZ (where N~$=$~Double, Triple, Quadruple) basis set corresponds to the uncontracted core-valence N-zeta \cite{Dyall:07,Dyall:12} Dyall's basis set for Bi and augmented correlation consistent polarized valence N-zeta, aug-cc-pVNZ \cite{Dunning:89,Kendall:92} basis set for light atoms.
In the DZC basis set the contracted version of the aug-cc-pVDZ \cite{Dunning:89,Kendall:92} basis sets were used for light atoms.

Based on the nonrelativistic estimates, the hybrid density functional PBE0 \cite{pbe0} has been chosen because it reproduces the nonrelativistic coupled cluster value rather well.
Geometry parameters of the BiF$_6^-$ anion have been obtained in the scalar-relativistic DFT calculation using the generalized relativistic pseudopotential method \cite{Mosyagin:16}.

The contribution of the Gaunt interaction to the shielding constant was estimated as the difference between  the values calculated at the Dirac-Hartree-Fock-Gaunt and Dirac-Hartree-Fock level of theory within the uncoupled scheme.

Nonrelativistic and scalar-relativistic calculations were performed within the {\sc us-gamess} \cite{USGAMESS1} and {\sc cfour} \cite{CFOUR} codes. Relativistic four-component calculations were performed within the {\sc dirac15} \cite{DIRAC15} and {\sc mrcc} \cite{MRCC2013} codes. For calculation of the hyperfine-interaction matrix elements and $g$ factors the code developed in Refs.\,\cite{Skripnikov:16b,Skripnikov:15b,Skripnikov:15a} was used.

\section{Results and discussion}

Table \ref{BiF6} contains results of the calculation of the BiF$_6^-$ anion.
\begin{table}[h]
\centering
\caption{The values of $^{209}$Bi shielding constants in BiF$_6^-$ in ppm.}
\label{BiF6}
\begin{tabular}{lccc}
\hline                    
\hline
Basis set/method     & Diamagnetic & Paramagnetic & Total \\
\hline                    
\hline
DZ-MB-LAO/DHF    & 8\,618 & 5\,768 & 14\,386  \\  

DZ-MB-LAO/DFT    & 8\,621 & 3\,726 & 12\,347  \\     
TZ-MB-LAO/DFT    & 8\,639 & 3\,733 & 12\,372  \\  
\hline    
DZC-RKB/DFT      &      & 3\,848    &    \\
DZC-RKB/CCSD     &      & 4\,403    &    \\
DZC-RKB/CCSD(T)  &      & 4\,286    &    \\
\hline    
QZ-MB-LAO/DFT    & 8\,628 & 3\,763 & 12\,391  \\  
Correlation correction &      &  437 &        \\  
Gaunt correction       &      &  -37 &        \\  
\hline                    
Final              &      &   & 12\,792  \\  
\hline                    
\hline                    
\end{tabular}
\end{table}
Comparing Dirac-Hartree-Fock (DHF) and DFT results in Table \ref{BiF6} it can be seen that the diamagnetic contribution to $\sigma(^{209}{\rm Bi})$ depends only weakly on the correlation effects, while the paramagnetic contribution is strongly affected.
To check the accuracy of the latter DFT result we have performed a series of relativistic coupled cluster calculations of $\sigma(^{209}{\rm Bi})$ taking into account only the positive energy spectrum. 
Comparing values obtained within the coupled cluster with single, double and noniterative triple-cluster amplitudes (CCSD(T)) with that of CCSD shows that the triple amplitudes only slightly contribute to $\sigma(^{209}{\rm Bi})$ demonstrating good convergence of the results with respect to the electron correlation treatment 
\cite{Note2}. 

In the final value of $\sigma(^{209}{\rm Bi})$ we include the correlation correction calculated as the difference between the CCSD(T) and PBE0 results.

To investigate the importance of systematic treatment of the  molecular environment 
 we have also performed an additional DHF study of one of the possible hydrated forms of Bi$^{3+}$ in an acidic solution of Bi(NO$_3$)$_3$ -- [Bi(H$_2$O)$_8$]$^{3+}$ cation in comparison with the unsolvated Bi$^{3+}$ cation.
It was found that the shielding constant of the $^{209}$Bi$^{3+}$ is significantly larger (by about 20\% at the DHF level) than that in [$^{209}$Bi(H$_2$O)$_8$]$^{3+}$.
Therefore, the interpretation of the \textit{molecular} NMR experiment in terms of the nuclear magnetic moment using a shielding constant obtained for the corresponding ion (as was done in earlier studies) is associated with considerable uncertainties.

We now use the value obtained for $\nu_\mathrm{^{209}BiF_6^-}/\nu_\mathrm{H}= 0.1607167(2)$ from our NMR measurements and the shielding constant of $\sigma(^{209}{\rm BiF}_6^-)=12\,792$\,ppm calculated above to obtain $\mu_I(\mathrm{^{209}Bi}) = 4.092(2)\,\mu_\mathrm{N}$ with an uncertainty dominated by theory. 

Table \ref{NewOld} compares the experimental values \cite{Ullmann:17} of the HFS splittings with the theoretical values calculated with the old [$\mu_I$(old)$=$4.1106(2)$\mu_N$] and the new [$\mu_I$(new)$=$4.092(2)$\mu_N$] values of the nuclear magnetic moment \cite{Volotka:12}. The theoretical results include the most elaborated calculation of the Bohr-Weisskopf effect \cite{Senkov:02}.
\begin{table}[]
\centering
\caption{Theoretical values of $\Delta E^{(1s)}$ and $\Delta E^{(2s)}$ (in meV) calculated with old and new nuclear magnetic moment of $^{209}$Bi in comparison with the experimental values \cite{Ullmann:17}.
For the Bohr-Weisskopf effect the most elaborated calculation by Sen'kov and Dmitriev \cite{Senkov:02} was employed.
}
\label{NewOld}
\begin{tabular}{llll}
\hline
\hline
   & \multicolumn{2}{l}{Theory}                             & Experiment         \\
      & $\mu_I$(old)  & $\mu_I$(new)      &    \\
\hline   
$\Delta E^{(1s)}$  & 5112(-5/+20)                & 5089(-5/+20)(2)  & 5085.03(2)(9)  \\
$\Delta E^{(2s)}$  & 801.9(-9/+34) & 798.3(-9/+34)(4) & 797.645(4)(14) \\          
\hline   
\hline
\end{tabular}
\end{table}
\begin{figure}[!h]
\includegraphics[width=0.98\linewidth]{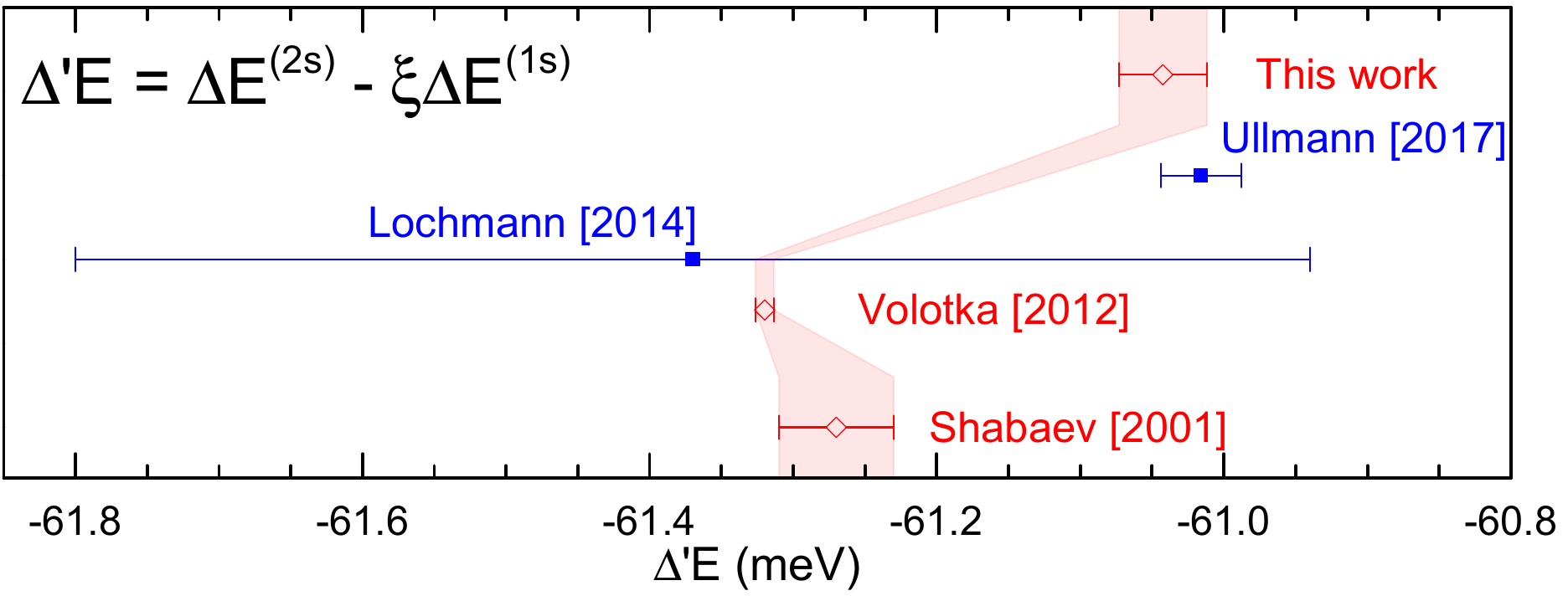}
\caption{\label{fig:CompExpTheory} Specific difference obtained in theory  \cite{Shabaev:01a,Volotka:12} (red) and experiment \cite{Lochmann:14,Ullmann:17} (blue). The new nuclear magnetic moment established in this work yields a new value for the specific difference which matches the most recent experimental value within uncertainty. 
}
\end{figure}
The new magnetic moment has been used to recalculate the specific difference and we obtain 
$\Delta 'E_{\mathrm{theo}} = -61.043(5)(30)$\,meV, where the first uncertainty is due to uncalculated terms and remaining nuclear effects, while the second one is due to the uncertainty of the nuclear magnetic moment obtained in the present work. Revised value of $\Delta 'E_{\mathrm{theo}}$ 
is plotted in Fig.\,\ref{fig:CompExpTheory} combined with the previous theoretical and experimental data. Theory and experiment are now in excellent agreement and the $7\sigma$ discrepancy reported in \cite{Ullmann:17} disappears. 
Unfortunately, the uncertainty of $\Delta 'E_{\mathrm{theo}}$ is now 14\% of the total QED contribution and about 1.5 times larger than the experimental uncertainty.
Hence, an improved value for the nuclear magnetic moment of $^{209}$Bi is urgently required, either from an atomic beam magnetic resonance experiment or from a measurement on trapped H-like ions. The latter will have the advantage that no shielding corrections have to be applied. Such an experiment is planned at the ARTEMIS trap \cite{Quint:2008} at the GSI Helmholtz Centre in Darmstadt. Only such a measurement combined with an improved determination of the HFS splitting in $^{209}$Bi$^{80+,82+}$ as it is foreseen at SPECTRAP \cite{Andelkovic:2013} can provide a QED test in the magnetic regime of strong-field QED. Our result also proves that a measurement of the specific difference can also be used to extract the nuclear magnetic moment. 
Doing so results in  $\mu_I(^{209}\mathrm{Bi})=4.0900(15)\,\mu_{\mathrm{N}}$ in excellent agreement with the NMR value obtained here. 

\begin{acknowledgments}
\section*{Acknowledgments}
We thank Petra Th\"orle from the Institute of Nuclear Chemistry at the University of Mainz for the preparation of the NMR samples and Dmitry Korolev from Saint-Peterburg State University for valuable discussions.
The development of the code for the computation of the matrix elements of the considered operators as well as the performance of all-electron coupled cluster calculations were funded by RFBR, according to Research Project No.~16-32-60013 mol\_a\_dk; performance of DFT calculations was supported by the President of Russian Federation Grant No. MK-2230.2018.2. This work was also supported by SPSU (Grants No. 11.38.237.2015 and No. 11.40.538.2017) and by SPSU-DFG (Grants No. 11.65.41.2017 and No. STO 346/5-1). The experimental part was supported by the Federal Ministry of Education and Research of Germany under Contract No 05P15RDFAA and the Helmholtz
International Center for FAIR (HIC for FAIR).   
\end{acknowledgments}


\newpage
\section{Supplementary: Synthesis of $\mathrm{(CH_3)_4N^+BiF_6^-}$}
 For the synthesis of $\mathrm{(CH_3)_4N^+BiF_6^-}$ (NMe$_4$BiF$_6$), all operations were carried out in an atmosphere of purified argon (4.8) in an all-metal vacuum line made out of stainless steel (SS 316). First, BiF$_5$ was synthesized by the direct fluorination of BiF$_3$ in a corundum boat with a mixture of F$_2$/Ar (50/50 V/V) at $450^{\circ}$C and subliming the product onto a corundum-coated monel cooling finger (water cooled, $12^{\circ}$C). 
 For the synthesis of NMe$_4$F, NMe$_4$OH was treated several
 times with aqueous HF and recrystallized from dry isopropanol
 \cite{Christe:90}.
 The synthesis of NMe$_4$BiF$_6$ was then carried out in a T-shaped FEP
 (fluorinated ethylene propylene) vessel. The educts, BiF$_5$ (129.7 mg, 0.43 mmol) and NMe$_4$F (39.9 mg, 0.43 mmol), decompose upon dissolution in aHF (anhydrous hydrogen fluoride) when they are in direct contact \cite{Morgan:83}. Therefore, in one leg of the FEP vessel  BiF$_5$ was suspended in aHF and in the other leg NMe$_4$F was dissolved in aHF. Both legs were cooled using liquid nitrogen to approximately $−80^{\circ}$C. The solution of NMe$_4$F was poured onto the BiF$_5$ suspension leading to the direct precipitation of colorless NMe$_4$BiF$_6$. The mixture was allowed to warm to room temperature, agitated for 5 minutes, and the solvent was removed in vacuo upon which the solid became yellow-orange. The yield was quantitative. The product was dissolved in acetonitrile to a saturated solution. The purity of NMe$_4$BiF$_6$ was evidenced using attenuated total reflection infrared (ATR-IR) and $^{19}$F-NMR-spectroscopy 
 at room temperature.

\end{document}